\documentclass[prb,preprint]{revtex4-1} 

\usepackage{amsmath}  
\usepackage{amsfonts} 
\usepackage{graphicx} 

\begin{document}


\title{Muon Bulb: A Cosmic-Ray Detector Inside a Light Bulb}

\author{Yuvaraj Elangovan}
\email{yue8@pitt.edu} 
 \affiliation{University of Pittsburgh} 



\date{\today}

\begin{abstract}
Cosmic-ray muons are among the most abundant high-energy particles reaching Earth's surface, with a flux of approximately one muon per square centimeter per minute at sea level, yet they remain entirely invisible to the naked eye, posing a persistent challenge for public engagement in particle physics. We present the Muon Bulb, a self-contained cosmic ray muon detector built within the enclosure of a standard commercial LED bulb, designed to produce a vivid and immediate visible light flash upon the passage of a cosmic-ray muon. The detector integrates a plastic scintillator tile coupled to a Silicon Photomultiplier (SiPM) for efficient detection of scintillation photons produced by ionizing muons. The SiPM signal is processed by a compact analog front-end comprising an amplifier and a leading-edge discriminator, which isolates genuine muon-induced pulses from background. Validated trigger signals are passed to an ESP32/RP2350 microcontroller, which drives an internal LED, causing the bulb to flash visibly in real time. The readout additionally enables wireless data transmission for real-time monitoring of muon event rates and flux measurements. The Muon Bulb is designed to serve as a high-impact outreach instrument across science museums, planetariums, K-12 classrooms and public science festivals without requiring specialized infrastructure. Its wireless capability enables networked classroom activities where multiple Muon Bulbs can be operated simultaneously, allowing students to collaboratively measure muon flux and engage in genuine data-driven scientific inquiry.  The Muon Bulb transforms a commercial house hold item into a powerful tool for particle physics, making the invisible universe visible to everyone.
\end{abstract}

\maketitle 

\section{Introduction}
Primary cosmic rays mostly protons and helium nuclei collide with nuclei in the upper atmosphere and initiate particle shower. The shower produces charged pions and kaons which decay into muons. A muon lives only $2.2\,\mu$s on average, but it travels so fast that time dilation stretches this out and lets it cross the atmosphere. Muons also lose energy slowly in air. For both reasons, muons are the most common charged particles that reach the ground.  The integral flux at sea level is close to one muon per square centimeter per minute and the angular distribution of the more energetic component is well described by $I(\theta) \propto \cos^{2}\theta$, where $\theta$ is the zenith angle \cite{pdg,grieder}. Every person is therefore traversed by a few hundred muons per second and yet nothing about ordinary experience suggests that this is happening.

\begin{figure}[h!]
\centering
\includegraphics[width=3in]{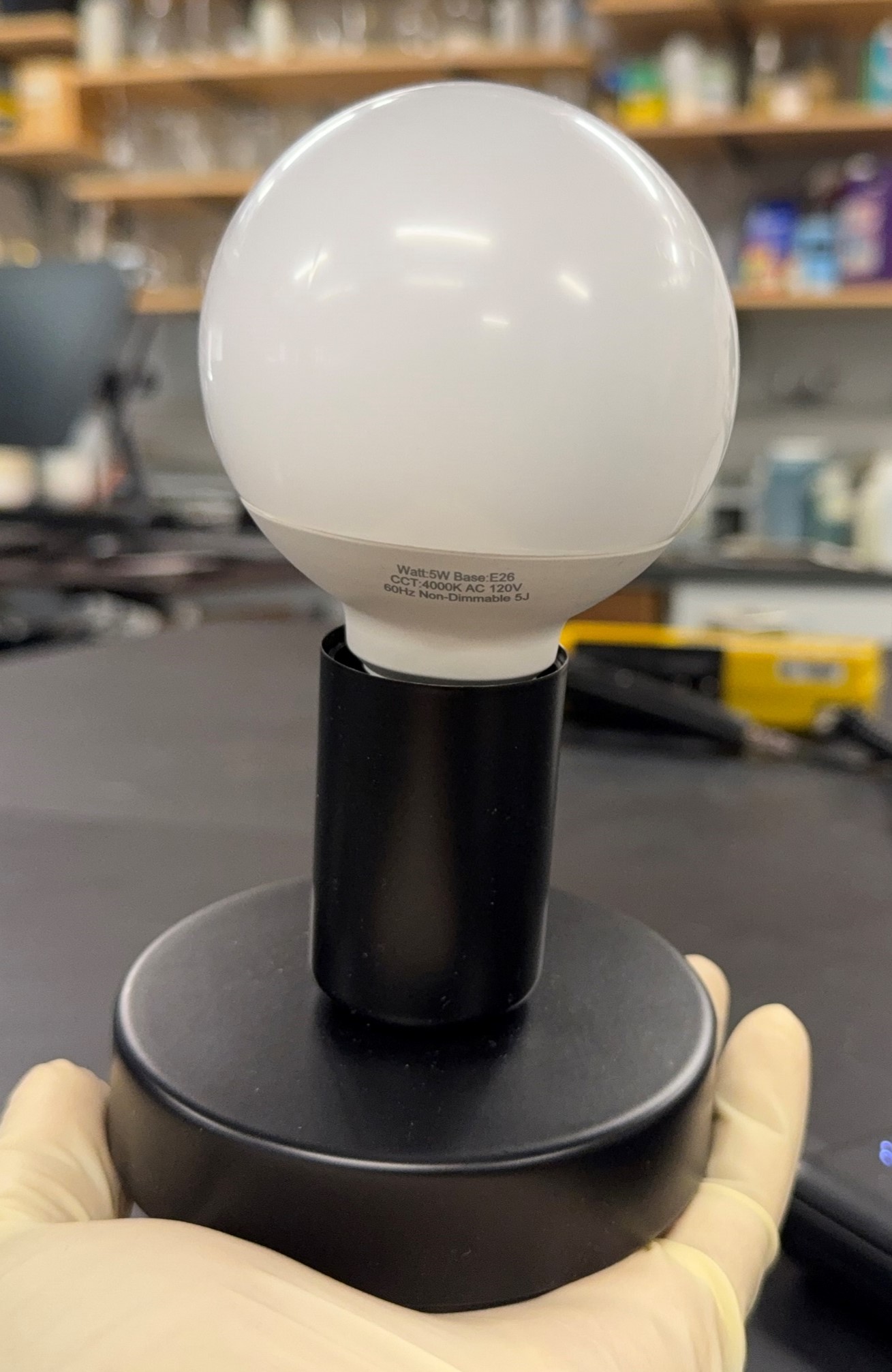}
\caption{The Muon Bulb Detector.}
\label{bulb}
\end{figure}

Muons are everywhere and yet nobody can see them. Converting these invisible traces of muons into visible signatures makes them good tool for particle physics outreach. Cloud chamber and Spark chamber gives a striking view of particle tracks, but it needs constant attention and Maintance time to time. A cosmic muon tracker based on Resistive Plate Chambers (RPCs) \cite{cmt} displays live muon tracks on an array of LEDs and is an excellent outreach tool, but it requires a continuous gas flow and someone on hand to operate it. for  A scintillator telescope read out on an oscilloscope or a laptop gives real numbers, but the audience has to be told which trace on the screen is a particle. Over the past ten years several small, low cost detectors built from plastic scintillator and silicon photomultipliers (SiPM) have appeared most notably the desktop muon detector and the CosmicWatch detectors that followed it \cite{dmd,cosmicwatch,cwphysics} and these have put real cosmic-ray measurements within reach of students and schools. The Muon Bulb shown in Figure.~\ref{bulb} takes a different approach to the display problem, instead of showing a number, it shows a light. The whole detector scintillators, SiPMs, front-end electronics, microcontroller and radio fits inside an ordinary LED bulb and the bulb flashes for one second every time a muon goes through it. A visitor sees a familiar light bulb object flashing at random and is told that each flash is one particle from space. We find that this one sentence explains more than any plot. The bulb also sends timestamped event data over a wireless link. This can be used to engage Students in Schools and universities in particle physics.

\section{WORKING PRINCIPLE OF THE Muon Bulb}

\begin{figure}[h!]
\centering
\includegraphics[width=4in]{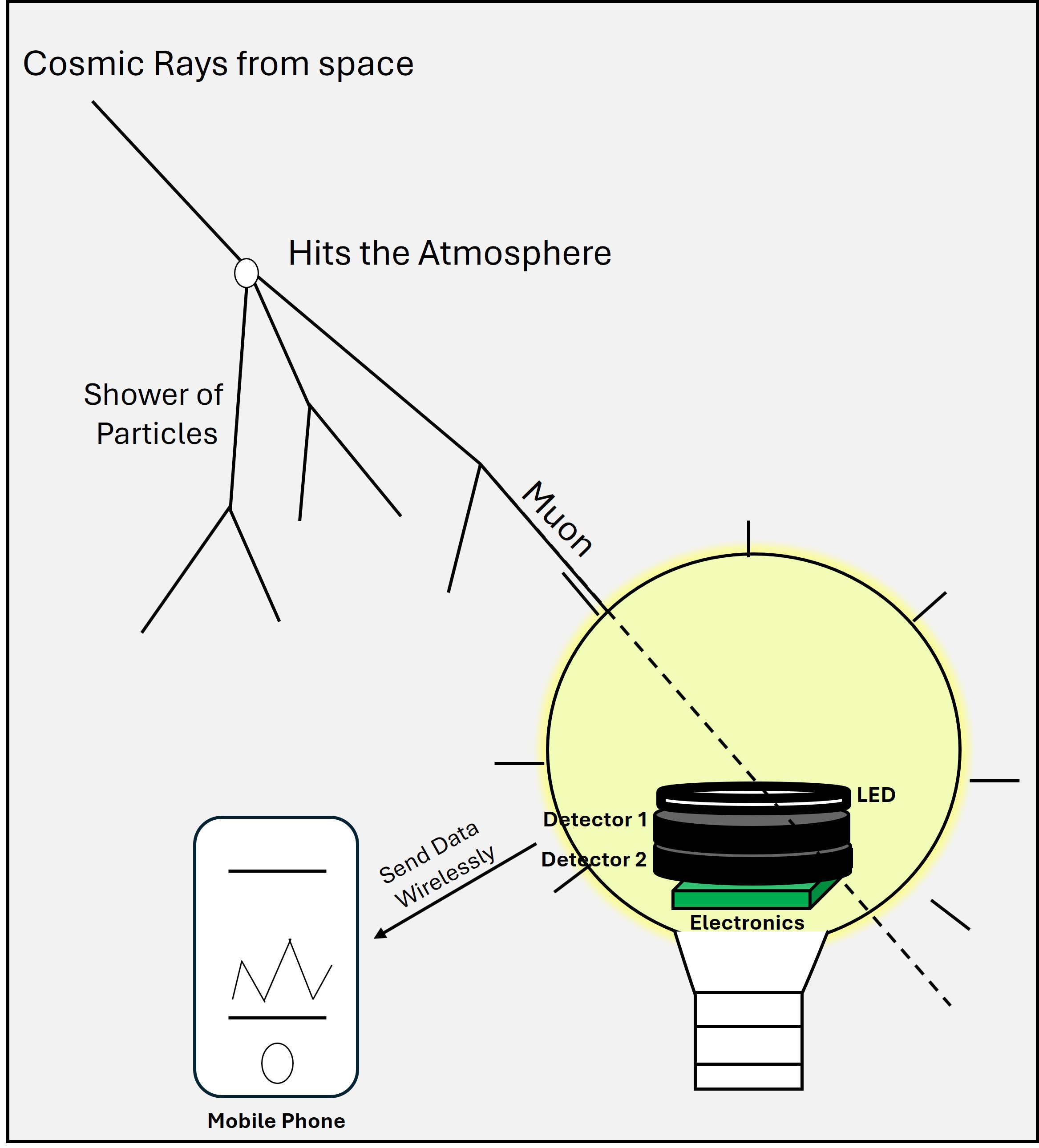}
\caption{Working Concept of the Muon Bulb Detector}
\label{concept}
\end{figure}

The Muon Bulb detects cosmic-ray muons by scintillation and shows each detection by lighting up an LED bulb, hence the name. The working concept of the muon bulb is shown in Figure.~\ref{concept}. A muon crossing the plastic scintillator \cite{ej200} detector is a minimum-ionizing particle, it loses energy in the plastic scintillator, exciting molecular states along its path. These states then de-excite by emitting photons. The light is turned into an electrical signal by a silicon photomultiplier (SiPM) \cite{sipm}, an array of tiny avalanche photodiode cells all biased a few volts above their breakdown voltage. When a photon is absorbed in one cell, that cell breaks down and produces a short pulse of charge that is always about the same size. All the cells share a single output, so the total signal is proportional to the number of cells that fired — in other words, to the light intensity. The SiPM signals are read out using off-the-shelf discrete electronics. To keep SiPM noise and background radiation out of the muon count, the detector uses two plastic scintillators stacked one above the other and looks for coincident signals from the pair. Each scintillator has its own readout channel, which converts the small analog pulse into a digital signal, and both digital signals are fed to an ESP32 microcontroller \cite{esp32}. When a valid coincidence occurs, the ESP32 lights the LED bulb for one second, producing a visible flash every time a particle passes through the detector. The SiPM bias voltage is kept constant so that the SiPM gain remains stable.

\section{Detector Geometry and Packing} 

\begin{figure}[h!]
\centering
\includegraphics[width=3in]{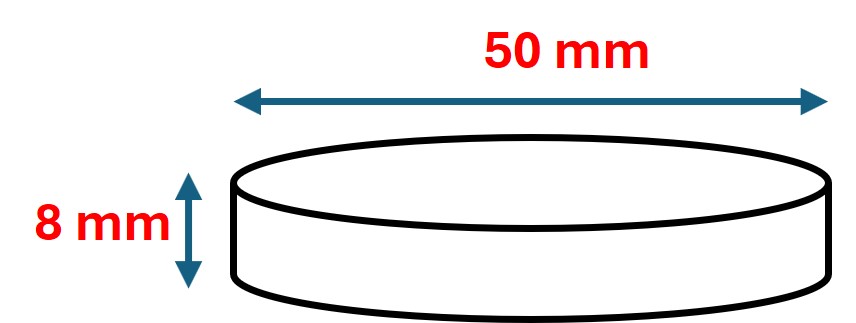}
\caption{Dimensions of the plastic scintillator disc}
\label{geo}
\end{figure}

The geometry is set by the bulb. A standard LED bulb has an internal envelope diameter ranging from $50\, \text{mm}$ to $100\, \text{mm}$. The detector, the readout board and the LED ring must all fit within the bulb circular area. We therefore use circular scintillator discs $50\,$mm in diameter and $8\,$mm thick, shown in Figure.~\ref{geo}, cut from standard EJ-200 and polished on all faces. Each disc consists of a volume of $15.7\,$cm$^{3}$. Around $20$ muons per minute pass through a single horizontal scintillator disc of this area.

\begin{figure}[h!]
\centering
\includegraphics[width=5in]{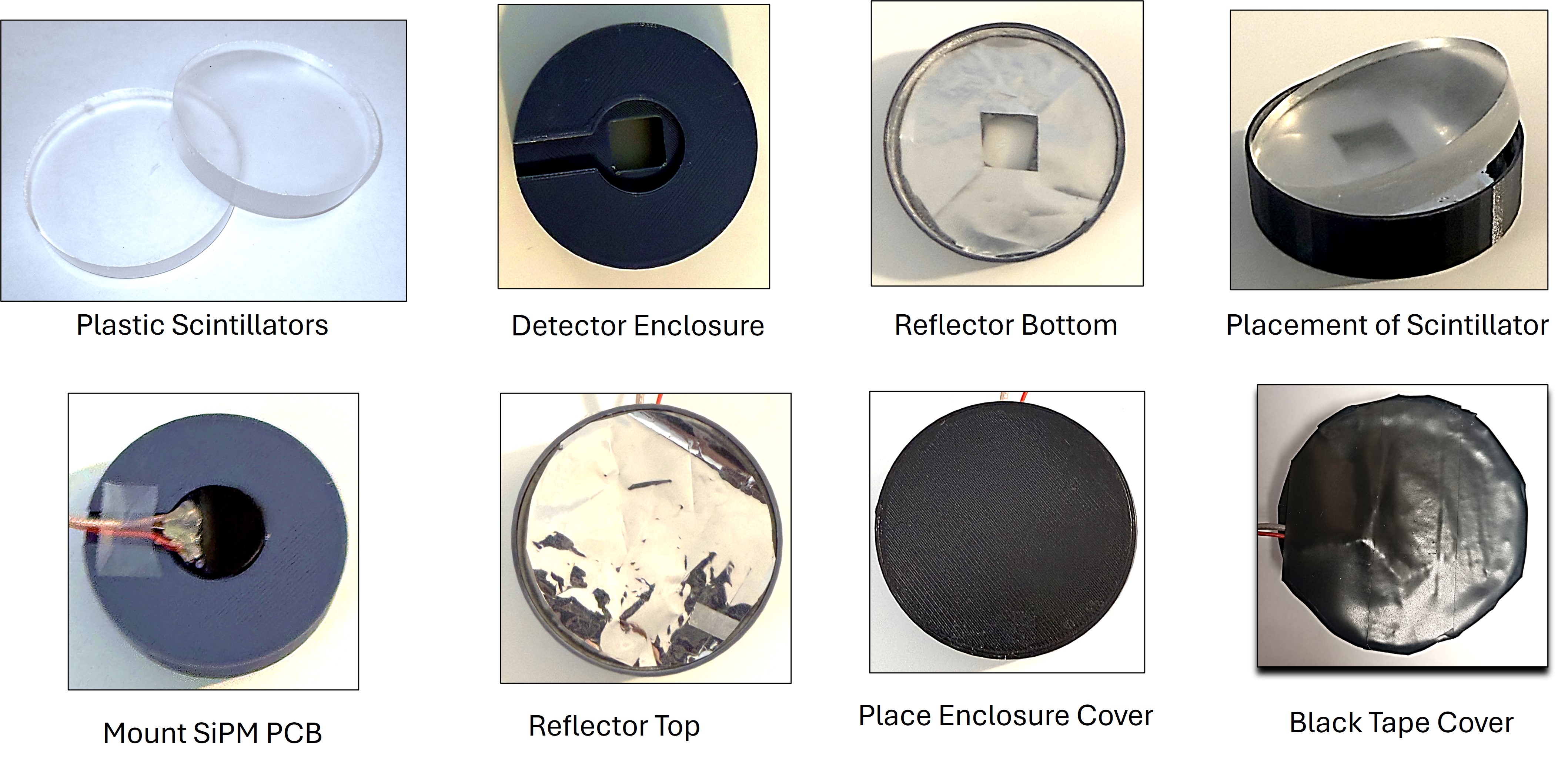}
\caption{Steps involved in packing a detector unit}
\label{det_pack}
\end{figure}

The scintillators are need to be packed tightly inside the light bulb. For better light collection by the SiPM all the faces of the scintillator must be covered with a reflector except the SiPM face.  The second is light tightness. A SiPM is a sensitive photon counter and the detector sits inside a bulb whose LED ring is bright, any leakage path from the ring to the sensor would affect the SiPM base line. So the detectors and SiPM are carefully assembled and packed to avoid leaks and to improve light yield. The packing sequence, shown in Figure.~\ref{det_pack}.

Each disc is first placed inside a 3D-printed cylindrical enclosure whose inner surface is lined with a reflector, The disc is seated on this reflector, a matching reflector is laid over its upper face and the SiPM carrier board is mounted at the center of of the scintillator using the given slot in the enclosure,  with the sensor facing the scintillator through thin air coupling. The enclosure cover is then fitted and the completed module is wrapped in black tape, giving a second independent light seal. The quality of the seal is verified before assembly by illuminating the closed module with a bright lamp and confirming that the single-channel rate does not change. Two identical modules are built and stacked on top with their flat faces sit together,

\section{Detector Assembly Inside a Bulb} 
\begin{figure}[h!]
\centering
\includegraphics[width=3in]{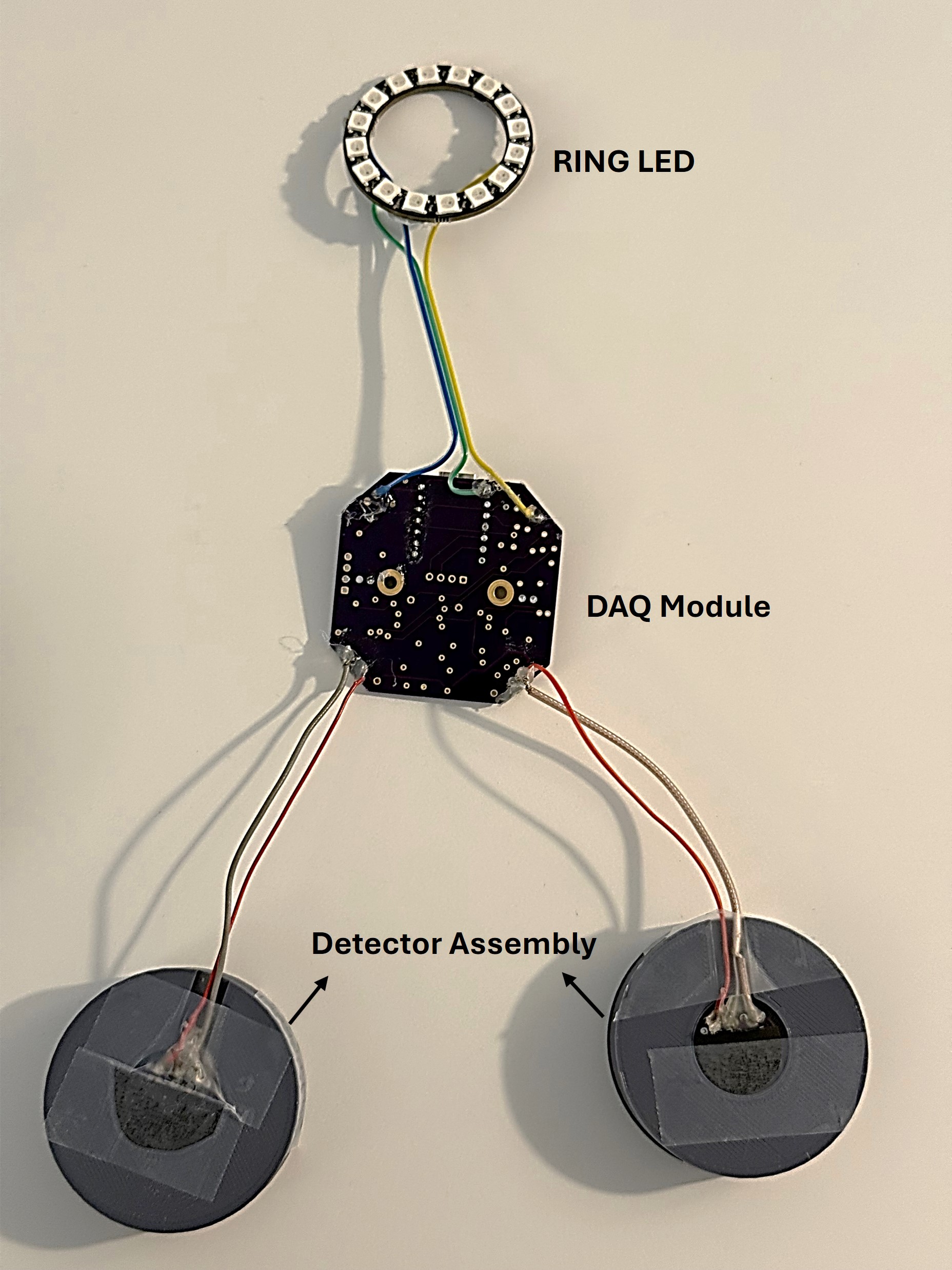}
\caption{Wired Detector Module}
\label{assembly}
\end{figure}

The two packed scintillator modules, the readout board and the display are integrated into a single mechanical unit, shown before insertion in Figure.~\ref{assembly}. The readout board carries the micro-controller, the two analog front ends and the SiPM bias generator and is connected to the two detector modules by short coaxial cables. An addressable RGB ring LED is mounted above the detector stack and driven from a single digital output of the micro-controller. The DAQ board is octagonal so that it clears the inner wall of the bulb base and its mounting holes are aligned with the bulb base holes which originally fitted with the LED bulb PCB. 

\begin{figure}[h!]
\centering
\includegraphics[width=6in]{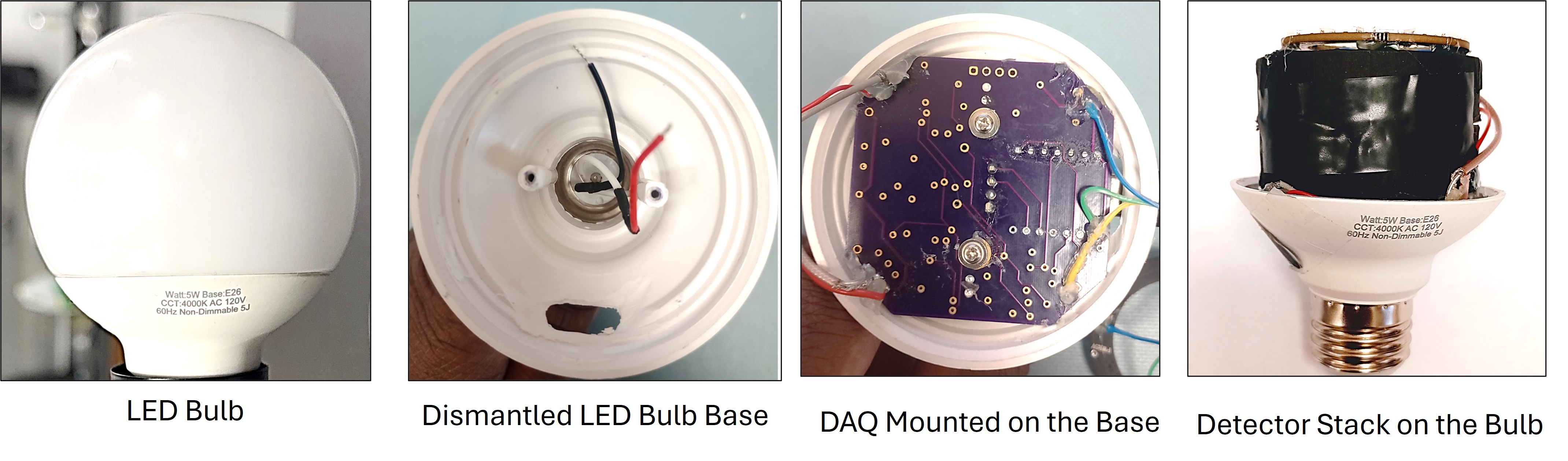}
\caption{Detector Assembly inside Bulb}
\label{bulbpack}
\end{figure}

Integration into the bulb is straightforward and requires very less machining. A commercial LED bulb is dismantled by disconnecting the diffuser from its base. Then the original driver board and LED array are removed from the base, leaving the moulded plastic base with screw shell as an empty housing. The screw shell is kept intact and for safety the instrument is never connected to mains voltage and is powered only from the 5\,V USB input. The readout board is fixed inside the base, the detector stack is placed on top of it, the LED ring sits above the stack and the original diffuser is refitted, so the finished object is externally indistinguishable from an ordinary bulb. Figure~\ref{bulbpack} shows these steps.

\section{DATA ACQUISITION SYSTEM}

\begin{figure}[h!]
\centering
\includegraphics[width=6in]{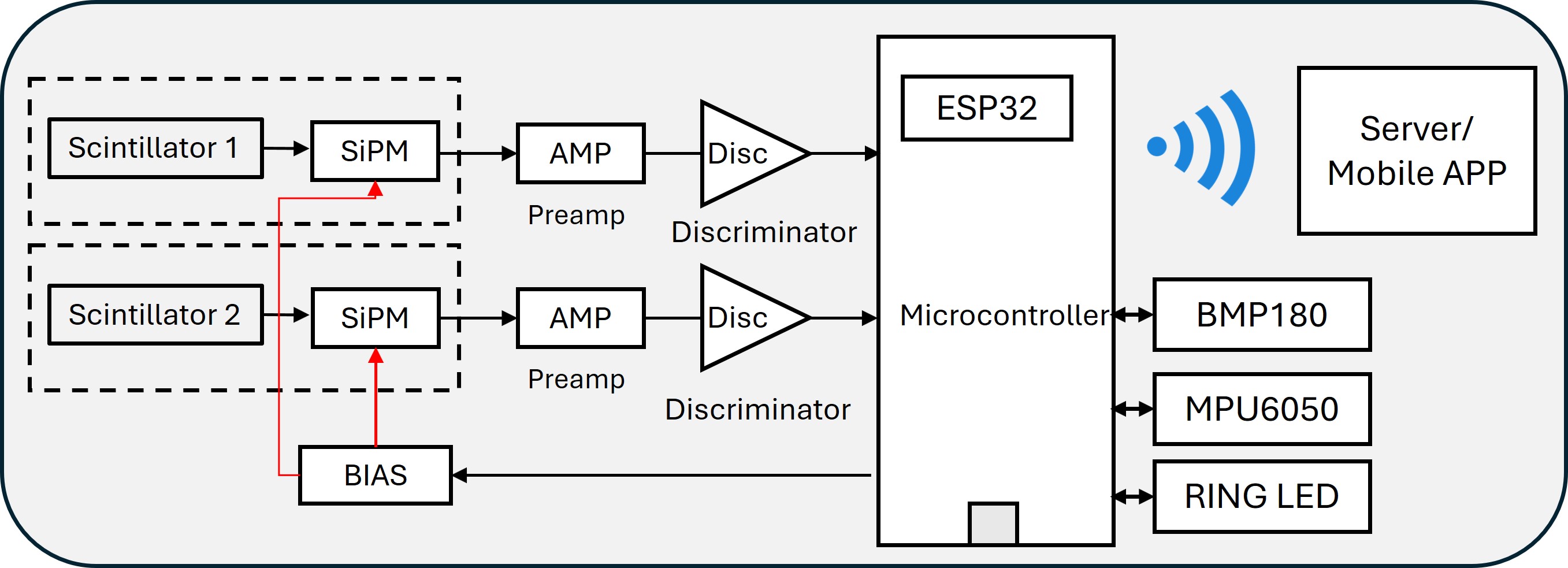}
\caption{Block Diagram of the Data Acquisition Module}
\label{daqblock}
\end{figure}

\begin{figure}[h!]
\centering
\includegraphics[width=4in]{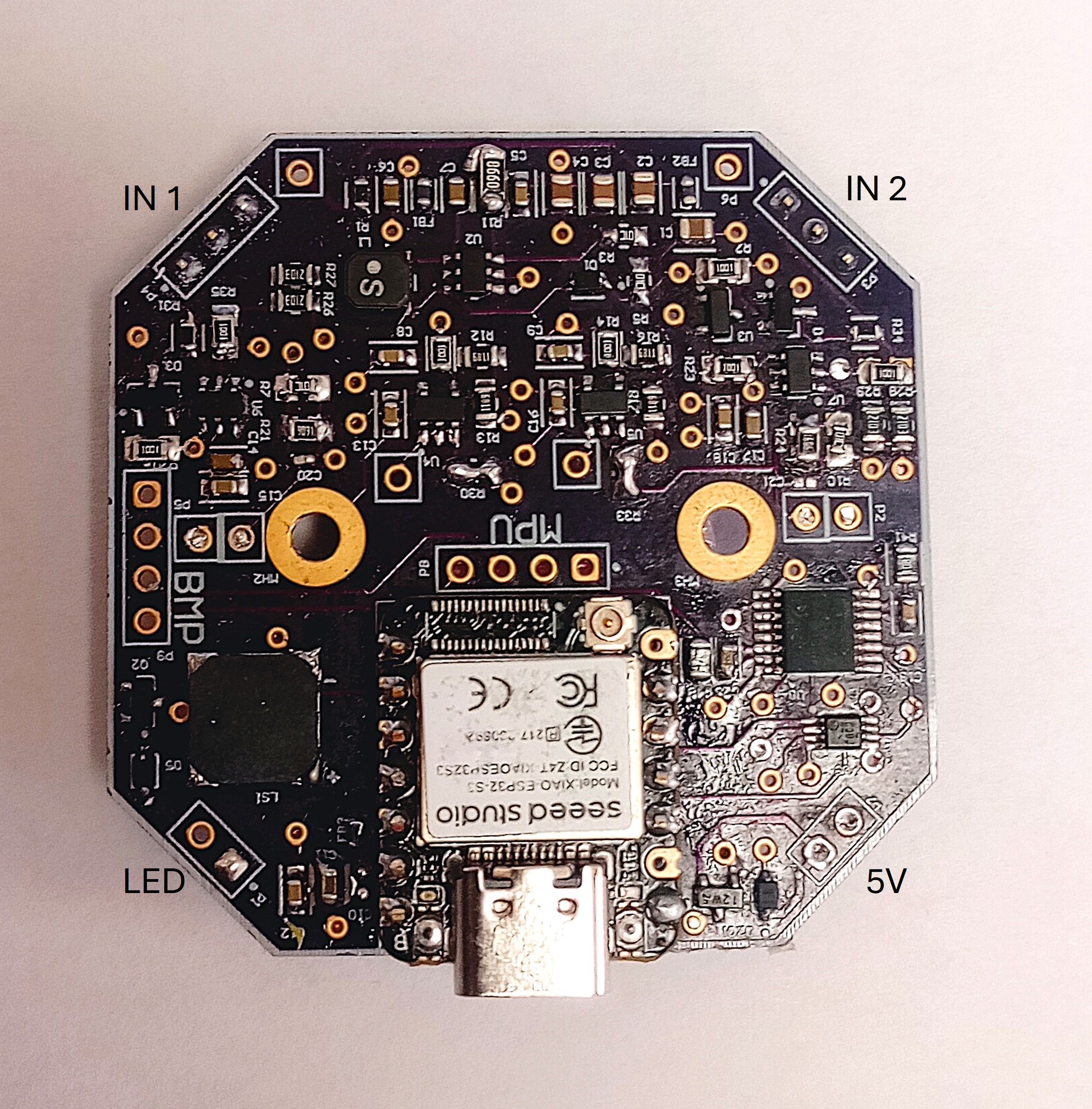}
\caption{DAQ PCB Module}
\label{daqpcb}
\end{figure}

The data acquisition system must do three things simultaneously: identify a coincidence event in real time; drive a bright LED without disturbing the analog front end and the SiPM; and record and transmit the event data (counts and temperature) wirelessly to a remote server. All three are handled by a single board, whose block diagram is shown in Figure.~\ref{daqblock} and whose physical realization is shown in Figure.~\ref{daqpcb}.

Each channel begins with the SiPM, whose fast output is terminated into an amplifier of voltage gain 20 and bandwidth 250 MHz. The amplified incoming pulses are compared against a fixed reference by a leading-edge discriminator, which produces a logic pulse. The threshold of around 200mV is set by a constant voltage source. This threshold level is a compromise between muon efficiency and noise and was fixed by measuring the singles rate as a function of threshold and choosing the point at which the steep dark noise contribution has flattened. Both SiPMs are supplied from a common bias generator on the same board, a boost converter followed by an RC filter that produces 30V from the $5\,$V rail. The bias is held constant rather than temperature compensated. 

The two discriminator outputs are routed to a Mono-shot which stretches the logic signals to 100 us. Then the signals are fed to an AND logic gate. This gate produces a Logic signal Trigger every time when there is a coincidence between both detectors. The trigger is routed to the interrupt-capable inputs of a Seed Studio XIAO ESP32-S3 module. As a redundancy both channels logic signals are also connected to the ESP32. On a coincidence trigger interrupt, the firmware writes the LED ring to full brightness for one second with red color. During normal running the ring is also used diagnostically: a hit in the upper tile alone lights the ring blue, a hit in the lower tile alone lights it green and a coincidence lights it red, as shown in Figure.~\ref{bulbdisplay}. This is a useful piece of teaching in itself, since an audience can watch the two single-tile colors flicker rapidly while the coincidence colour appears only occasionally and can thereby be shown rather than told why the coincidence requirement is necessary.
 
The firmware runs a programmable timer that accumulates event information over a 10-second interval and transmits the data at the end of each interval. Each record contains the counts from the two individual detectors and from the coincidence. A BMP180 sensor provides barometric pressure and temperature and an MPU6050 accelerometer supplies the orientation of the bulb, so that the run log records whether the detector was vertical or horizontal. Pressure and temperature are logged because the muon rate at ground level depends measurably on both.

\begin{figure}[h!]
\centering
\includegraphics[width=5in]{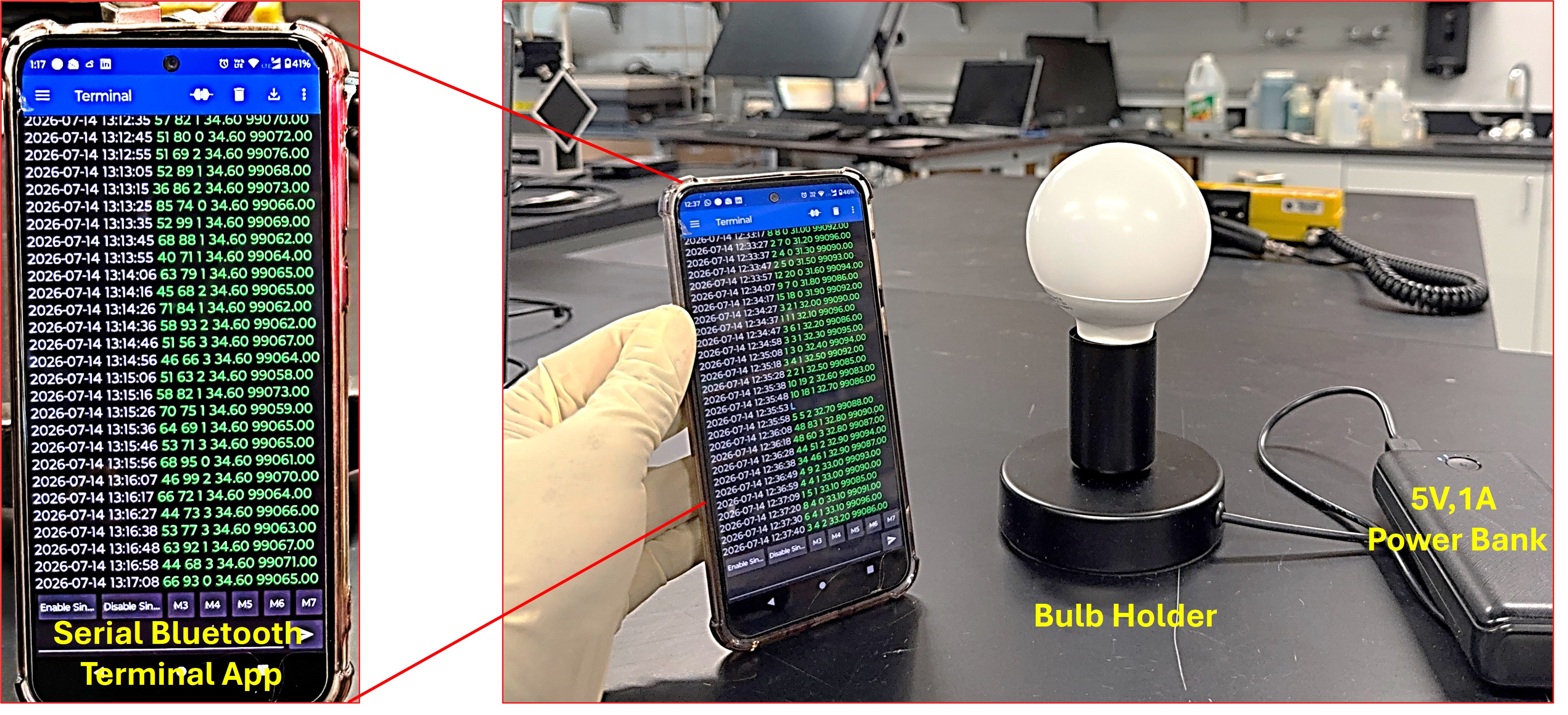}
\caption{Serial Bluetooth Terminal logging Muon Bulb Data over Bluetooth}
\label{app}
\end{figure}

The ESP32-S3 supports both Bluetooth and Wi-Fi. In the simplest mode the microcontroller transmits the data in Bluetooth and Bluetooth terminal application on a phone displays the event stream as it arrives. Figure~\ref{app} shows the bulb running on a $5\,$V, $1\,$A power bank with its data being read on a phone Android app. This mode needs no infrastructure and no installation, which is what makes it practical in a museum or a classroom. In the networked mode the micro-controller joins a local access point and posts each event to a web server. so that several bulbs can be logged together and their rates compared.

\section{Results}
We evaluated the Muon Bulb in three ways. First, we verified that the display behaves as intended and remains stable over the long. Second, we compared the coincidence rate of the bulb in vertical and horizontal positions, which tests whether the instrument really is responding to the cosmic-ray flux and not to something local. Third, we expose the detector to a radioactive source to confirm that both channels respond to ionizing radiation.

\subsection{LED Bulb Display}
\begin{figure}[h!]
\centering
\includegraphics[width=4in]{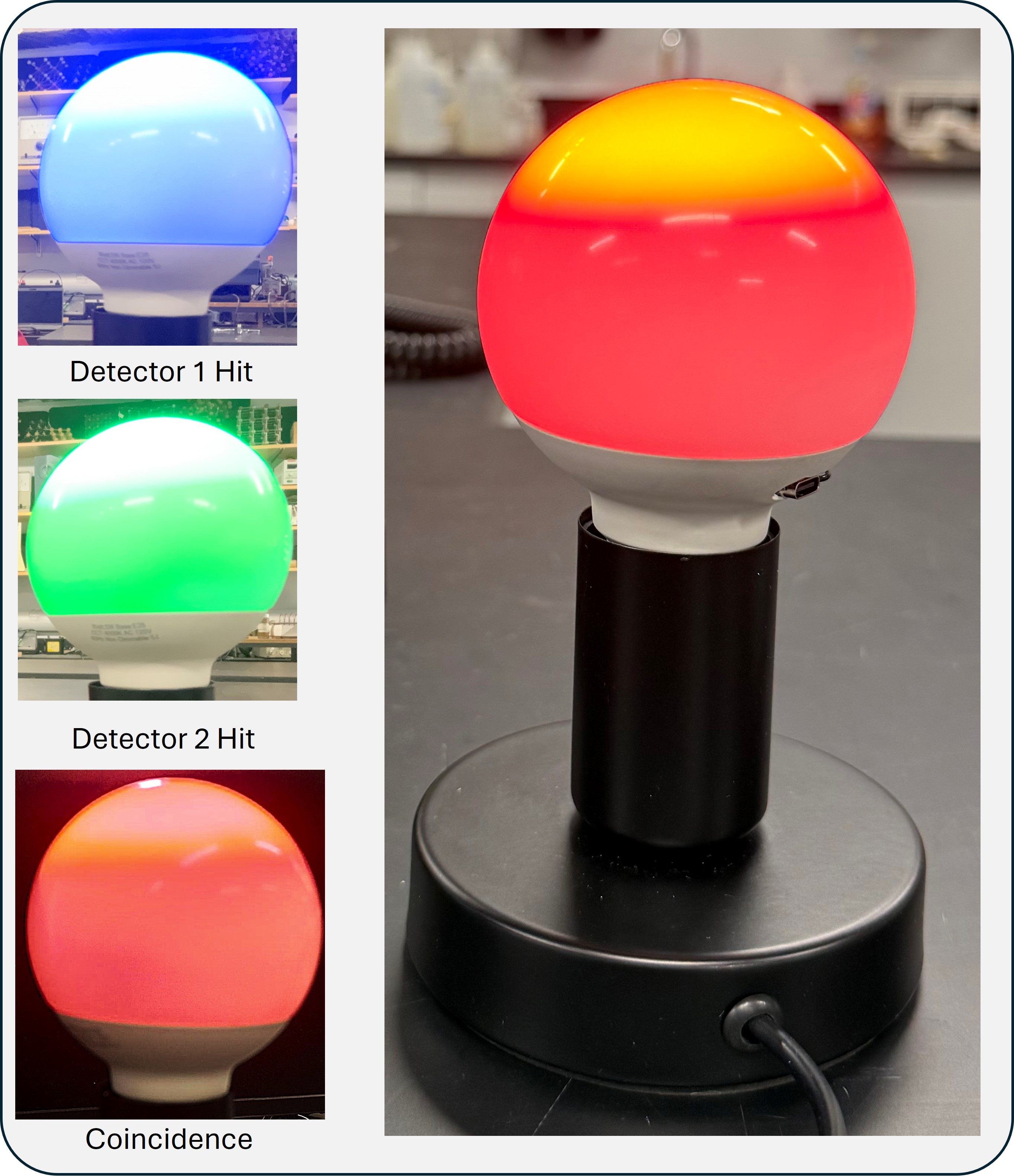}
\caption{Muon Bulb Visualization}
\label{bulbdisplay}
\end{figure}

Figure~\ref{bulbdisplay} shows the bulb during operation. The three panels on the left show the three display states: a hit in the first detector alone, a hit in the second detector alone and a coincidence. The large panel on the right shows a coincidence flash as an observer sees it, with the entire diffuser illuminated. The flash is visible in normal room lighting and at a distance of several meters, 

The display was run continuously overnight without intervention. During that period the coincidence rate was stable to within few \,Hz and no failures of the LED ring, the micro-controller or the wireless link were observed. When the bulb ran continuously in a room with no air conditioning, its temperature rose slightly. In an air-conditioned room, the temperature stayed nearly constant and no effect in coincidence measurement.

\subsection{Horizontal Vs Vertical Flux}
\begin{figure}[h!]
\centering
\includegraphics[width=4in]{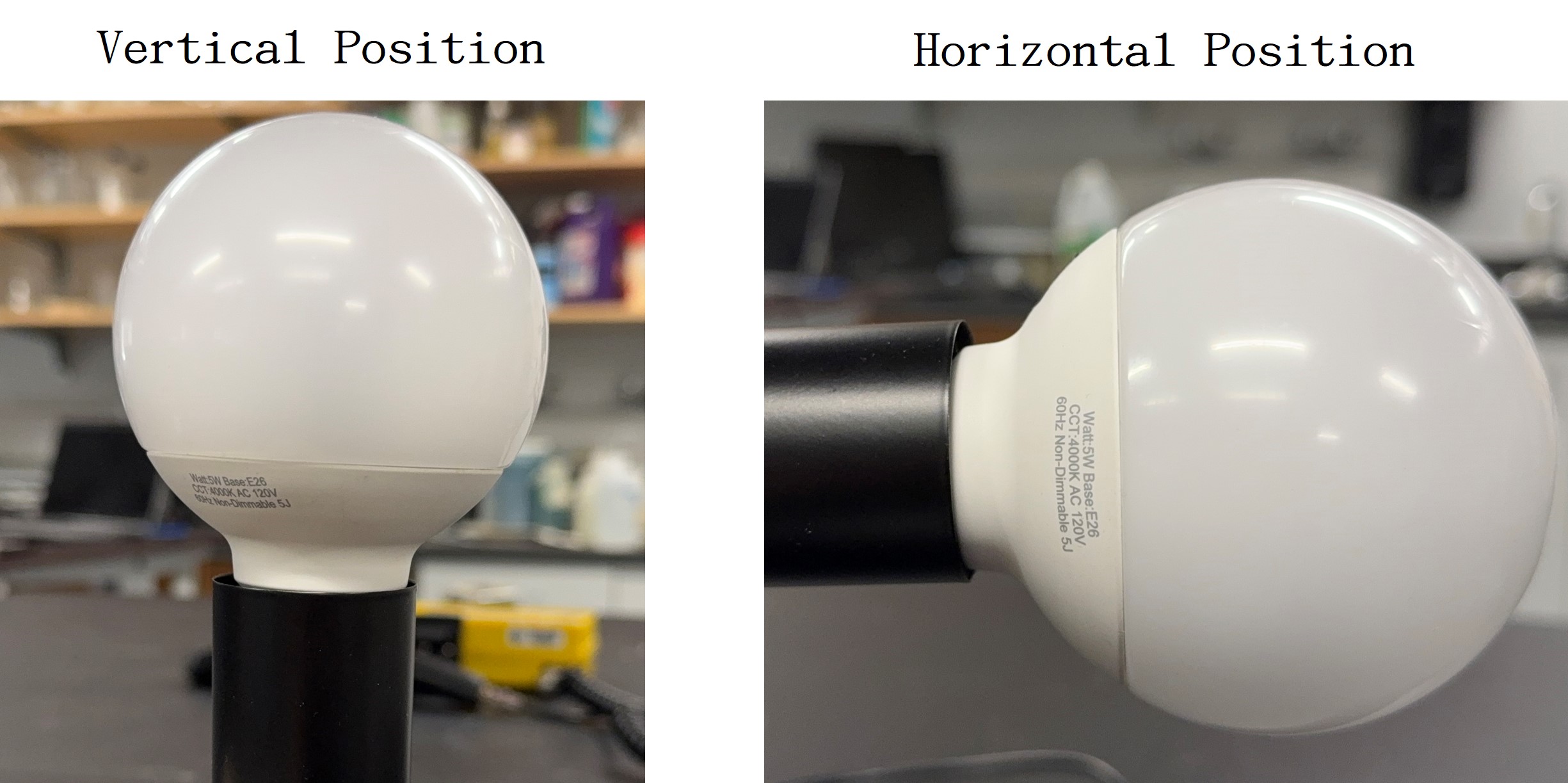}
\caption{Measurement Orientation of the Muon Bulb}
\label{fluxpos}
\end{figure}

To check the response of the muon bulb to cosmic muon flux, the bulb was tested in two different configuration Horizontal and vertical as shown in Figure~\ref{fluxpos} In the vertical position the scintillator planes are horizontal and the telescope axis is vertical; in the horizontal position the planes are vertical and the axis is horizontal.

\begin{figure}[h!]
\centering
\includegraphics[width=5in]{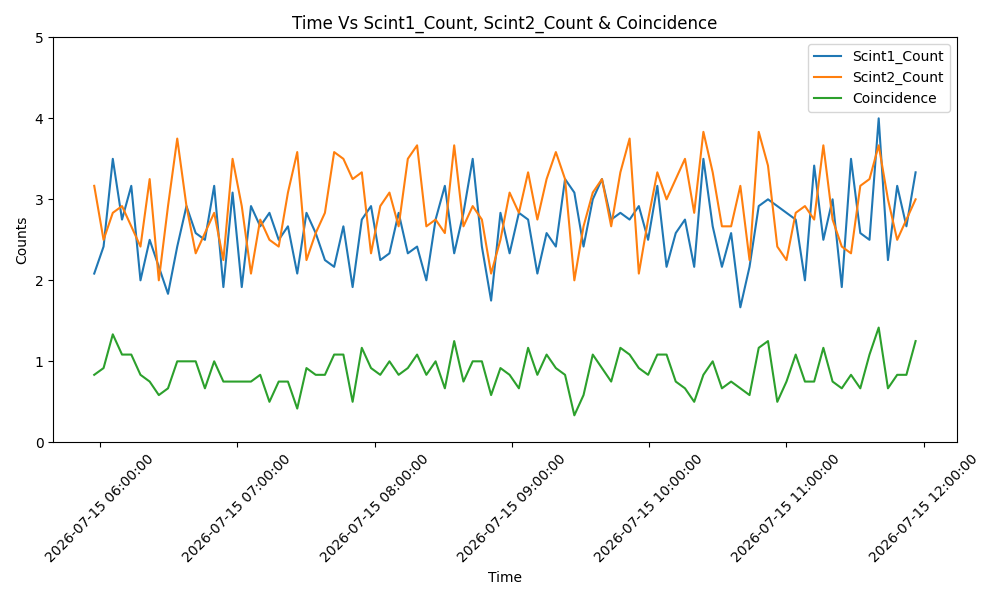}
\caption{Vertical Muon Flux Vs Time}
\label{vertflux}
\end{figure}
 
\begin{figure}[h!]
\centering
\includegraphics[width=5in]{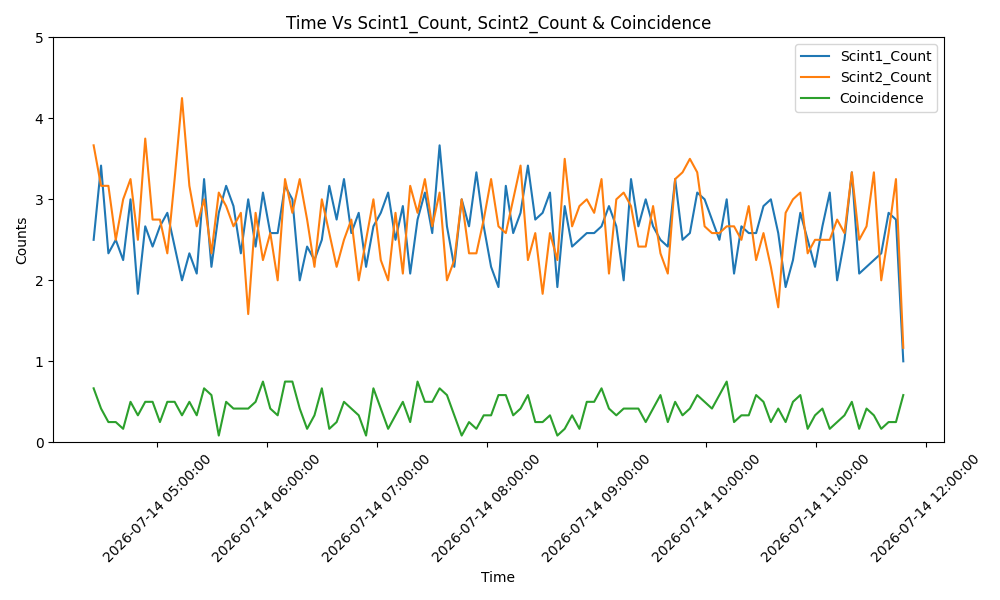}
\caption{Horizontal Muon Flux Vs Time}
\label{horzflux}
\end{figure}

Figures~\ref{vertflux} and~\ref{horzflux} show the two single-channel rates and the coincidence rate as a function of time for runs of $\sim 6hrs$ in the vertical and horizontal orientations respectively, taken at the surface. In both orientations the singles rate of the two channels are consistent with each other and stable. The coincidence rate, in contrast, is clearly higher in the vertical orientation.

That a household bulb can resolve the angular dependence \cite{source} of the cosmic-ray flux is, we think, the single most useful measurement it makes for teaching purposes: it requires nothing but tipping the instrument over, it takes a few hours of data and it produces a result that cannot be explained by any local effect.

\subsection{Source Testing of the Muon Bulb}
 \begin{figure}[h!]
\centering
\includegraphics[width=4in]{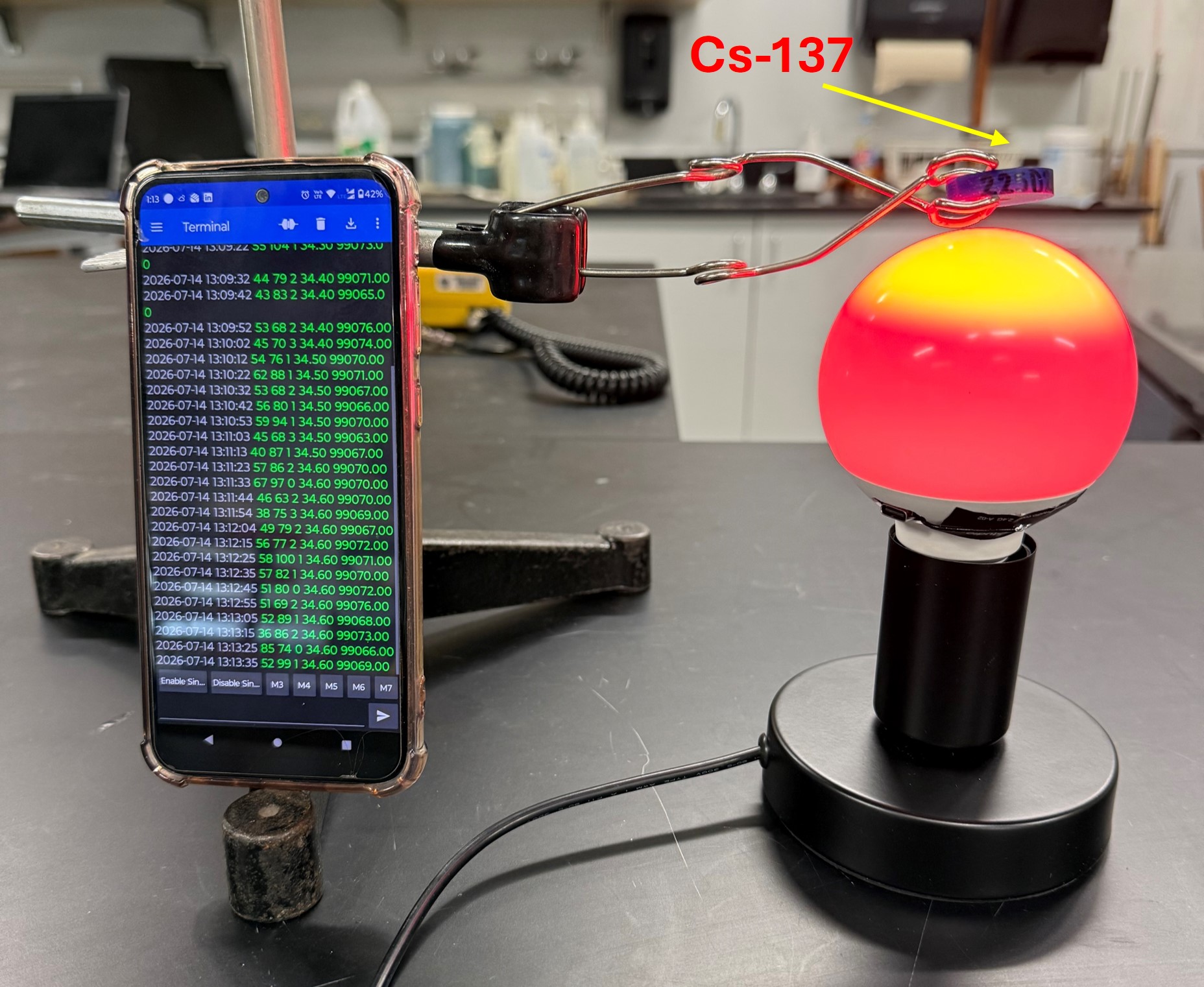}
\caption{Source Testing setup.}
\label{sourcetest}
\end{figure}
\begin{figure}[h!]
\centering
\includegraphics[width=5in]{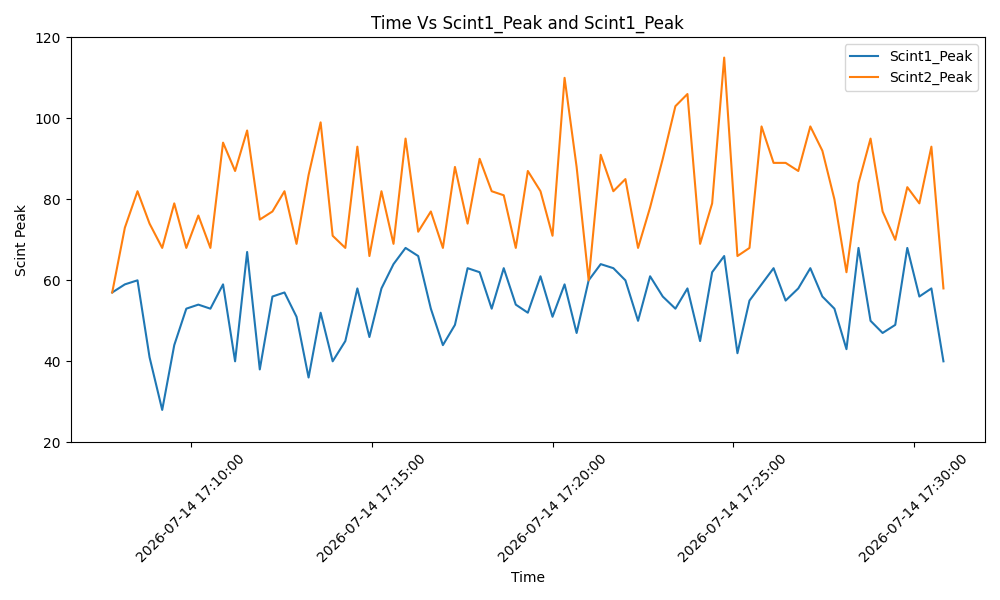}
\caption{Scintillator Counts Vs Time with Cs-137 Source kept at 1\,cm above bulb}
\label{sourceplot}
\end{figure}

To confirm that the response of both channels is due to ionizing radiation,the bulb was exposed to a gamma source placed $\sim 1\,cm$ above the bulb as shown in Figure~\ref{sourcetest}. We noticed the bulb was blinking faster than usual \cite{source} and the channels count rates were high. The single-channel rate rises from few\,Hz to $\sim 60\,Hz$ shown Figure~\ref{sourceplot}. in when the source is introduced and returns to its original value when it is removed, confirming that the response is due to the source and not to an electronic artifact.

\subsection{OUTREACH}
The Muon Bulb has been used in multiple venues for outreach, most recently at the DPF 2026 meeting. Three things have been consistent across these settings. First, the object requires no advertisement: visitors approach a flashing bulb without being invited, which is not true of an oscilloscope trace. Second, the randomness of the flashes is itself the teaching moment, audiences quickly ask why the flashes are irregular, which opens directly onto counting statistics and the Poisson distribution and older students can be handed the logged data and asked to test it. Third, the two-color singles display makes the coincidence argument concrete: visitors can see that the individual detectors fire constantly and that coincidence between them is rare and so understand without any mathematics why two detectors are better than one.
 
Concepts that are ordinarily difficult in a short interactions like particle trajectories, the atmospheric origin of secondary muons, the reason a detector must reject its own noise and the angular dependence of the flux can be demonstrated on the using the muon bulb during outreach \cite{bulb} . 


\section{Future Scope and Expansion}
\begin{figure}[h!]
\centering
\includegraphics[width=5in]{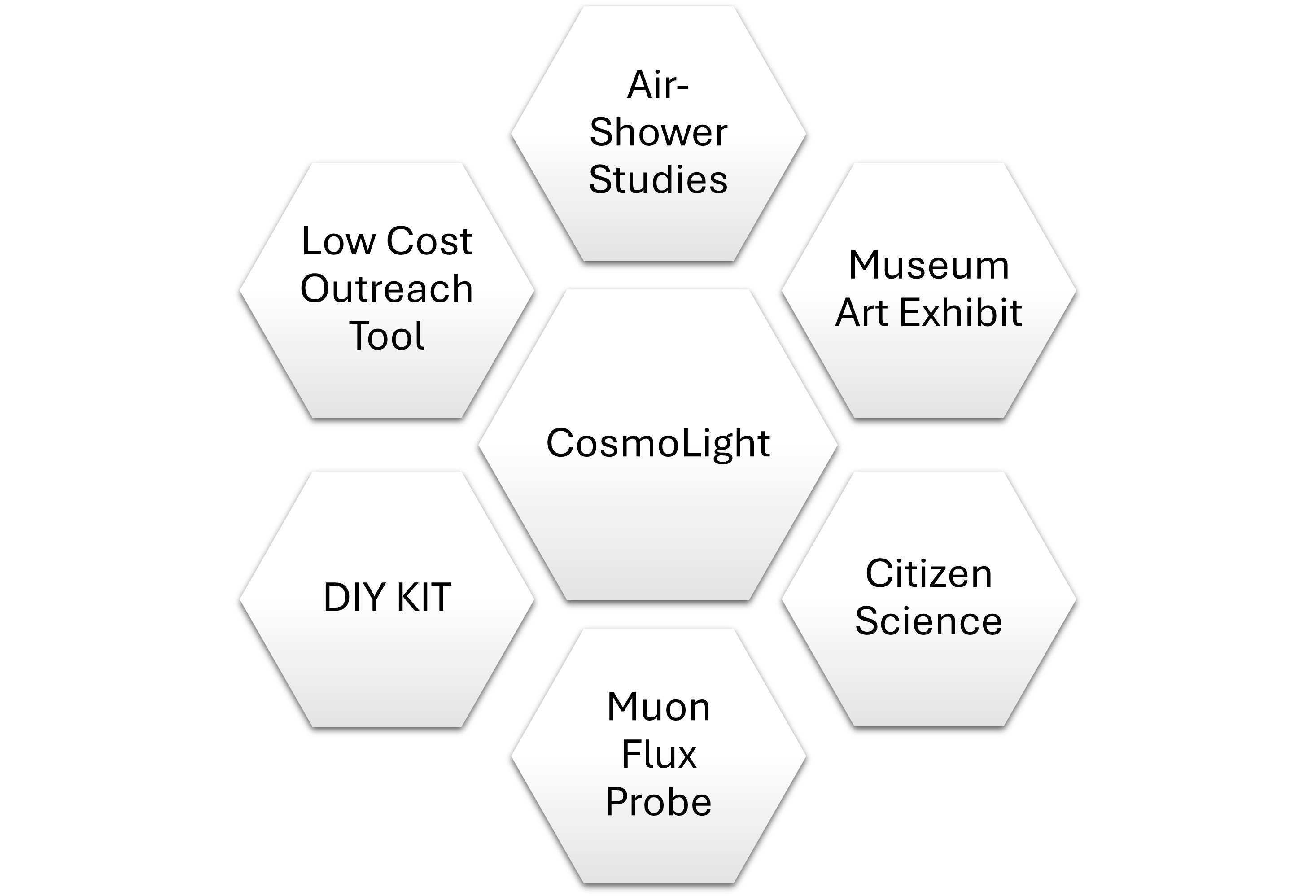}
\caption{CosmoLight Goals}
\label{cosmolight}
\end{figure}

The bulb was designed as a single self-contained unit, As an extension of this work we are creating a nation wide program called CosmoLight. The objectives of CosmoLight is to educate, engage students and public with particle physics concepts. Figure~\ref{cosmolight} summaries the directions we are pursuing under the name CosmoLight. One of our goal is to engage high school students and under graduate students in building the muon bulb and contributing to CosmoLight physics goals. 

\begin{figure}[h!]
\centering
\includegraphics[width=5in]{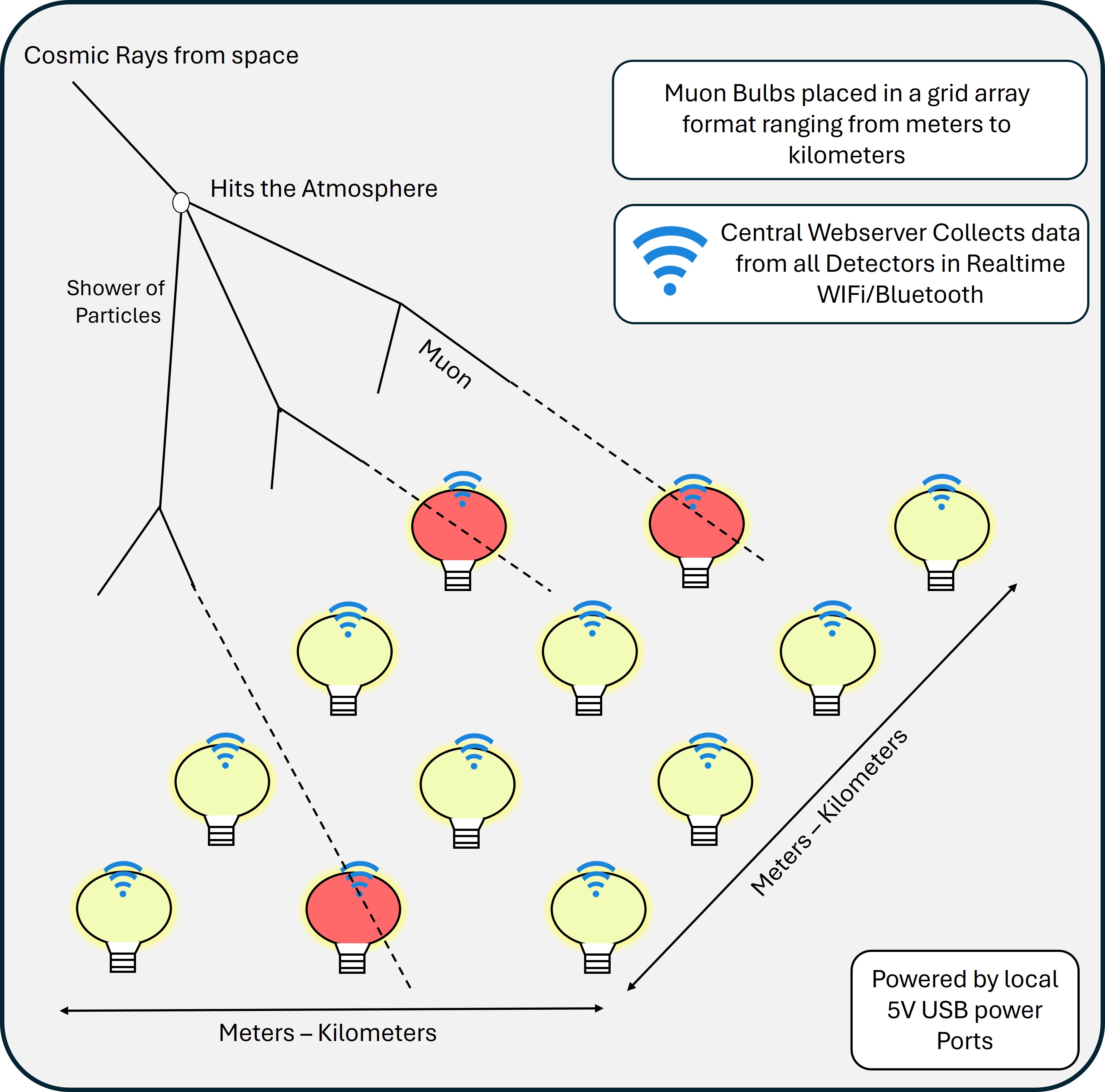}
\caption{Muon Bulb Detector Array Concept}
\label{array}
\end{figure}

Also a  distributed array of bulbs, separated by meters to kilometers and reporting to a common server as sketched in Figure.~\ref{array}, would allow the search for time-correlated hits in widely separated detectors that signal an extensive air shower. This measurement requires only that each unit to be time-synchronised better than few\,ms, which is achievable over the network. Grouping many bulbs into a single sculpture or installation turns the same hardware into a public art piece that flickers with real particles. Adding sound and brightness control extends this further. Because the coincidence rate falls steeply with overburden, carrying a bulb into a tunnel, cave, mine, or subway station gives a direct measurement of muon attenuation. Finally, because the bill of materials is short and the assembly requires no machining, distributing Do-It-Yourself (DIY) kits with a build guide supports  citizen science campaigns in which distributed volunteers contribute real cosmic-ray data. Currently multiple versions of muon bulb are under production. one of them did not need any micro-controller working only with hardware coincidence circuit. Also plans to fabricate 100 muon bulbs as a pilot program is underway. 

\section{Conclusion}
The Muon Bulb is a complete cosmic-ray muon detector inside an ordinary light bulb.
Its performance is adequate for real measurements as well as for display. The coincidence rate is stable over runs of several hours and it falls by a factor of half when the bulb is turned from the vertical to the horizontal orientation, in agreement with the expected $\cos^{2}\theta$ zenith-angle dependence of the sea-level muon flux. 
Because pressure, temperature and orientation are logged with every event, longer runs give access to the barometric dependence of the rate. An easy plug and play approach makes muon bulb to be used in different geo locations without large infrastructure.  By putting a working particle detector inside an object that everyone already knows how to interpret, the Muon Bulb makes a continuous, invisible flux of particles from space into something a visitor can simply watch happen.




 \begin{acknowledgments}
 Work performed in Dietrich School Electronics Shop Core Facility (RRID:SCR 025113) and services and instruments used in this project were graciously supported, in part, by the University of Pittsburgh. The author is supported by the National Science Foundation (award no. PHY-1948993). I thank my advisors Tae Min Hong and Pranava Teja Surukuchi for their guidance and encouragement in the outreach initiatives at the Department of Physics and Astronomy. I am grateful to the Electronics Shop and the Machine Shop of the Dietrich School of Arts and Sciences for supporting this development with tools and materials, and to Richard Misura for his assistance with the radioactive source testing.



 \end{acknowledgments}



\begin{thebibliography}{99}

\bibitem{pdg} R. L. Workman \textit{et al.} (Particle Data Group), ``Review of Particle Physics,'' Phys. Rev. D \textbf{110}, 030001 (2024).
 
\bibitem{grieder} P. K. F. Grieder, \textit{Cosmic Rays at Earth: Researcher's Reference Manual and Data Book} (Elsevier, Amsterdam, 2001).

\bibitem{cmt} Yuvaraj Elangovan \textit{et al.} ``Design and development of portable RPC-based Cosmic Muon Tracker,'' Journal of Instrumentation, \textbf{20}, P09004 (2025), 

\url{https://doi.org/10.1088/1748-0221/20/09/P09004}.

\bibitem{dmd} S. N. Axani, J. M. Conrad, and C. Kirby, ``The desktop muon detector: A simple, physics-motivated machine- and electronics-shop project for university students,'' Am. J. Phys. \textbf{85}, 948--958 (2017), \url{https://doi.org/10.1119/1.5003806}.
 
\bibitem{cosmicwatch} S. N. Axani, K. Frankiewicz, and J. M. Conrad, ``The CosmicWatch Desktop Muon Detector: a self-contained, pocket sized particle detector,'' J. Instrum. \textbf{13}, P03019 (2018).
 
\bibitem{cwphysics} S. N. Axani, ``The physics behind the CosmicWatch desktop muon detectors,'' arXiv:1908.00146 (2019).

\bibitem{ej200} \textit{EJ-200 Plastic Scintillator Data Sheet}, Eljen Technology. 

\url{https://eljentechnology.com/products/plastic-scintillators/ej-200-ej-204-ej-208-ej-212}.
 
\bibitem{sipm} \textit{Silicon Photomultiplier (SiPM) MICROFC-60035-SMT-TR Datasheet}, onsemi. 

\url{https://www.onsemi.com/pdf/datasheet/microc-series-d.pdf}.
 
\bibitem{esp32} \textit{XIAO ESP32 S3 Module}, Seed Studio.

\url{https://www.seeedstudio.com/XIAO-ESP32S3-p-5627.html?srsltid=AfmBOorpraeiQ7VU9XT-7p8IsQjNI2JCb7_ra82170c1jhyhmRBJyHON}.

\bibitem{expo} Yuvaraj Elangovan, Shashwat Kakkad,  and B. Satyanarayana, ``Cosmic Muon Explorer: a portable detector for cosmic muon flux measurements and outreach,'' Physics Education, \textbf{61}, 035006 (2026), \url{https://doi.org/10.1088/1361-6552/ae5371}.

\bibitem{source} Video of a Source testing of the Muon Bulb, 

\url{https://youtu.be/3uSfv0KhA08}.

\bibitem{bulb} Video of a live muon event in the Muon Bulb, 

\url{https://www.youtube.com/shorts/bfYkHWVfqHM?feature=share}.




\end{thebibliography}
\end{document}